# Deep-learning-enabled inverse design of large-scale metasurfaces with full-wave accuracy

Borui Xu[1], Jingzhu Shao[1], Xiangyu Zhao[1], Haishan Xu[2], Yudong Tian[1], Nanxi Chen[3], Jielin Sun[4], Han Lin[5], Qiaoliang Bao[6], Yiyong Mai[2,*], and Chongzhao Wu[1,*]

[1]Center for Biophotonics, Institute of Medical Robotics, School of Biomedical Engineering, Shanghai Jiao Tong University, Shanghai 200240, China

[2]State Key Laboratory of Synergistic Chem-Bio Synthesis, School of Chemistry and Chemical Engineering, Shanghai Jiao Tong University, Shanghai 200240, China

[3]Shanghai Institute of Microsystem and Information Technology, Chinese Academy of Sciences, Shanghai, China and University of Chinese Academy of Sciences, Beijing, China

[4]Institute of Translational Medicine, Shanghai Jiao Tong University, Shanghai 200240, China

[5]School of Science, Royal Melbourne Institute of Technology University, Melbourne, Victoria 3000, Australia

[6]Institute of Energy Materials Science (IEMS), University of Shanghai for Science and Technology, Shanghai 200093, China

mai@sjtu.edu.cn and czwu@sjtu.edu.cn

**Abstract:** Recent advances in meta-optics have enabled diverse functionalities in compact optical devices; however, conventional forward design approaches become inadequate as device complexity and scale grow. Inverse design offers a powerful alternative but often requires massive computational resources and neglects mutual coupling effects. Here, we propose and experimentally validate a deep-learning-enabled framework for rapid inverse design of large-scale, aperiodic metasurfaces with full-wave accuracy. The framework integrates an inverse design network responsible that maps target near-field responses to metasurface geometries in a non-iterative and scalable manner. A lightweight forward prediction network, integrated as a full-wave solver surrogate within the framework, enables efficient end-to-end training of the inverse design network while capturing mutual coupling effects by considering both local and neighboring geometries. The framework's effectiveness is experimentally verified through a multi-foci metalens and a holographic metasurface. This framework enables the inverse design from micrometer to centimeter scales ($> 20\text{k}\lambda$), with near-field responses discrepancies less than 3% compared to full-wave solvers at subwavelength ($< \lambda/10$) resolution. Moreover, it is generalizable to metasurfaces of arbitrary size and operates efficiently without high-performance resources, overcoming the computational bottlenecks of previous inverse design methods.

## 1. Introduction

Meta-optics, a revolutionary category of flat optical devices, has transformed the methodology to manipulate light and electromagnetic (EM) waves by leveraging the precise modulation capability of the tiny, called "meta-atoms"[1,2]. Recent advances in physical principles[3–5] and large-scale meta-optics manufacturing techniques[6,7] indicate that meta-optics may play a ubiquitous role in advanced photonic applications. Researchers have showcased numerous innovative fields that can be largely improved by meta-optics, including ultra-compact imaging cameras[8,9], augmented reality systems[10], and miniaturized spectrometers[11]. When it comes to practical applications, the precise design of large-scale or highly intricate meta-optical systems is highly required. For example, optical systems with large objective apertures are crucial for collecting weak or rapidly varying signals, as in astronomical imaging[12,13] and aerial surveillance[14]. They are also indispensable in high-power laser systems[15], where larger apertures help reduce incident power density and prevent material damage. In such a context, most of current meta-optics rely on a conventional "forward" methodology, where individual meta-atoms are manually tuned to match a precalculated phase distribution[16]. Although forward design approaches can be effective for simple tasks, their reliance on numerous preconceptions limits creative optimization and becomes inadequate for large-scale or highly intricate meta-optical systems. As design problems grow in complexity, scale, or imposed constraints, the effectiveness of forward design methods in identifying optimal solutions diminishes significantly. To unlock the full potential of meta-optics, a radically new design framework is urgently desired.

In contrast to forward design, inverse design begins with predefined functional objectives and employs computational algorithms to iteratively optimize structural geometries. This paradigm has emerged as a powerful framework for addressing large-scale, complex engineering problems, such as the optimization of aerodynamic structures or mechanical systems. In the field of photonics, inverse design has recently gained significance as a revolutionary methodology[17–22]. By harnessing computational optimization to explore non-intuitive and unconventional geometries beyond empirical design intuition, inverse design enables superior device performance and multifunctionality that often surpasses the capabilities of traditional forward design approaches. Several variants of inverse design methodologies have been explored. One major category is topological optimization, which employs gradient-based algorithms to iteratively optimize photonics structures[23–32]. Another approach involves deep-learning based techniques[33–36], where neural networks are trained either to directly map device geometries to target responses or to generate high-performance designs through generative models such as generative adversarial networks.

However, inverse design of large-scale aperiodic meta-optics remains constrained by limited accuracy and prohibitive cost of forward simulations. Topological optimization requires multiple iterations of forward simulations, while deep-learning methods relay on high-fidelity EM data generated from forward simulations for training. As design dimensions grow, high-accuracy forward simulations become intractable due to the multiscale nature of the problem—spanning from nanoscale meta-atoms to macroscale meta-optics. It is unrealistic to simulate an aperiodic metasurface with a 1-cm diameter using finite element method (FEM) or Finite difference time domain (FDTD). These EM simulation solvers are utilized to provide rigorous nanoscale physical accuracy, but as a result, they face prohibitive computational demands for full-scale device modeling. For instance, simulating a metasurface of 50 μm$^2$ size (assuming a 5-nm mesh size) by a FDTD solver typically demands ~100 hours of computation time and 100 GB of memory resources[23].

To alleviate the computational burden, some recent studies have adopted simplified modeling strategies by assuming periodic boundary conditions around each meta-atom, effectively treating aperiodic metasurfaces as locally periodic structures[37–39]. This approximation enables efficient forward prediction and inverse design by reducing the complexity of the EM environment surrounding each meta-atom. Nevertheless, such periodic assumptions fail to capture the mutual coupling effects present in realistic aperiodic configurations. To address this issue, deep-learning-enabled forward predictors have been developed to explicitly model coupling effects, which indeed improves accuracy but still suffers from relatively low efficiency. As a result, studies on large-scale metasurfaces based on these coupling-aware forward predictors are still limited to micrometer-scale configurations, which are not applicable to centimeter-scale metasurfaces[28–31]. The representative state-of-the-art deep-learning-enabled inverse design methods for metasurfaces reported over the past decade are summarized in Table S1 of the Supporting Information, highlighting the lack of scalable solutions for large-area, aperiodic designs.

In this work, we present a deep-learning-enabled framework for the rapid inverse design of large-scale, aperiodic metasurfaces with full-wave accuracy. Within this framework, an inverse design network directly maps target near-field responses to metasurface geometries in a non-iterative and scalable manner. To enable end-to-end training of the inverse design network, a forward prediction network is first trained to rapidly and accurately estimate near-field EM responses, and subsequently integrated as a fixed full-wave solver surrogate for metasurfaces of arbitrary size. To explicitly account for mutual coupling effects, the forward prediction network predicts the local response of each target meta-atom based on its geometry and that of its neighboring meta-atoms[30–32], as illustrated in **Figure 1**. By replacing conventional full-wave solvers with a learned surrogate model, the framework substantially reduces computational cost, enabling the inverse design of macroscale ($> 1k\lambda$) metasurface in tens of seconds on a standard four-core CPU without sacrificing full-wave accuracy. More importantly, it supports inverse design across scales ranging from micrometers to centimeters ($> 20k\lambda$), corresponding to an area expansion of five orders of magnitude compared with comparable studies[28–30], while still achieving discrepancies less than 3% compared to high-accuracy FDTD simulations performed at subwavelength ($<\lambda/10$) resolution. This approach achieves efficient performance for both forward prediction and inverse design, thereby overcoming the scale limitations and computational inefficiencies of conventional methods, broadening the scope of metasurface design to previously inaccessible regimes. To the best of our knowledge, this is the first demonstration of rapid inverse design for large-scale metasurfaces with full-wave accuracy, paving the way for scalable, efficient, and physically consistent metasurface design in practical optical systems.

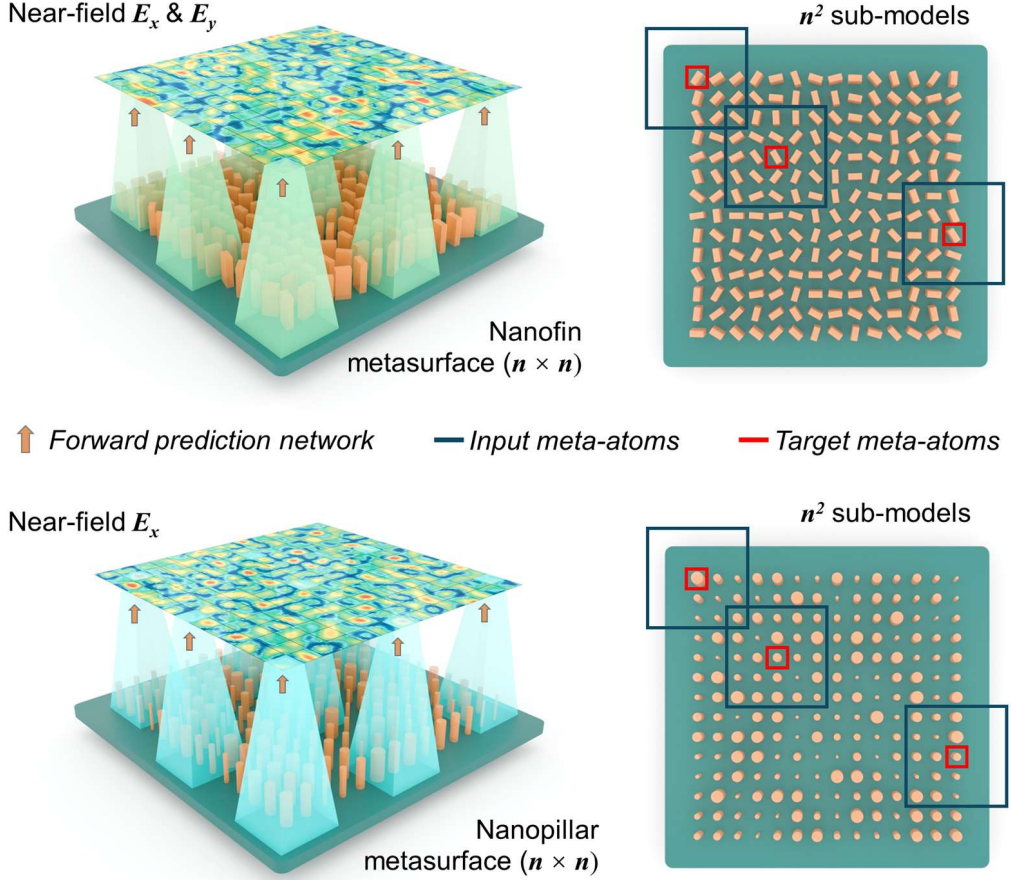

**Figure 1.** Schematic of forward prediction for near-field EM responses incorporating mutual coupling effects. The $n \times n$ dimension metasurfaces, comprising PB-phase-based nanofins (top) and propagation-phase-based nanopillars (bottom), can be decomposed into $n^2$ sub-models (highlighted by the blue box), each containing a target meta-atom (highlighted by the red box) and its neighboring meta-atoms as input to the forward prediction network.

## 2. Mutual Coupling Effects and Dataset Generation

Without loss of generality, we employ a transmissive all-dielectric metasurface architecture based on TiO$_2$ nanostructures fabricated on a SiO$_2$ substrate[1,9]. The platform integrates two distinct functional elements: (a) rectangular TiO$_2$ nanofins for Pancharatnam-Berry (PB) phase modulation and (b) cylindrical TiO$_2$ nanopillars for propagation phase control, as illustrated in Figure 1. The designed metasurfaces operate at a wavelength of 532 nm under left-handed circular polarization (LCP) for nanofin configuration and at a wavelength of 500 nm under *x*-polarization for nanopillar configuration. Both configurations are under normal incidence and with a uniform lattice constant $p = 400$ nm. The nanofins feature a height of 500 nm and in-plane dimension of 125 nm × 275 nm, while the nanopillars feature 500 nm tall with radius ranging from 25 to 125 nm. By carefully selecting the geometric parameters of nanofins and nanopillars, the tailored meta-atom library can achieve full $2\pi$ phase coverage without compromising transmission efficiency—a critical requirement for diverse optical applications.

To obtain accurate local response of randomly generated meta-atoms surrounded by diverse neighboring environments (hereinafter referred to as "local responses"), a quantitative analysis of mutual coupling effects between adjacent meta-atoms was conducted. First, the periodic-boundary response of a single meta-atom was computed via FDTD simulations and designated as the target meta-atom. Eight meta-atoms were then randomly selected from the meta-atom library and arranged around the target meta-atom to form a 3 × 3 array, referred as the 1-layer model. By analogy, additional neighboring meta-atoms were successively added to construct $N$-layer models ($N$ = 1, 2...,10), while periodic-boundary model was defined as 0-layer model. The local responses of the target meta-atom in each model were monitored under perfectly matched layers (PMLs), except for 0-layer model, as illustrated in **Figure 2**a-b. To generate the dataset for this analysis, 100 distinct 10-layer models were initially constructed. From each 10-layer model, a series of reduced $N$-layer models ($N$ = 9, 8, ..., 0) were sequentially derived by progressively removing the outermost neighboring layers. The average mean square errors (MSEs) of amplitude and phase between the $N$-layer and ($N$–1)-layer models were computed as a quantitative measure of their difference, defined as Equation 1-2:

$$MSE_{ampl} = \frac{\sum_{ij}\left(A_{N_{ij}} - A_{N-1_{ij}}\right)^2}{10 \cdot 10} \qquad (1)$$

$$MSE_{phase} = \frac{\sum_{ij}(abs(P_{N_{ij}}) - abs(P_{N-1_{ij}}))^2}{10 \cdot 10} \qquad (2)$$

where $A_N$ and $P_N$ are the local amplitude and phase responses of the target meta-atom in the $N$-layer model, sampled using a 10 × 10 array of point monitors, as shown in the inset of Figure 2c-d. The sampling resolution is 40 nm ($< \frac{1}{10}\lambda$).

As shown in Figure 2c-d, the simulation results of target meta-atoms under periodic boundary conditions exhibit significant discrepancies in both amplitude and phase compared to those of 1-layer model, with MSEs of (0.38, 0.67 rad) and (0.56, 0.86 rad) respectively. These results highlight the strong influence of mutual coupling effects. The local response of a target meta-atom can be significantly influenced by even 1-layer neighboring meta-atoms. As the number of neighboring layers $N$ increases, the MSE gradually decreases. However, when $N$ > 5, the MSE drops to less than 5% of its peak value, indicating diminishing variation. This trend is also visually evident in the amplitude and phase distributions of local responses, as shown in Figure 2e–f, confirming the reduced impact of additional neighboring layers. To balance accuracy and computational cost, $N$ = 5 was chosen for data generation. The final dataset contains the geometric parameters and local responses of over 10,000 randomly generated 5-layer meta-atom models. For nanofins, the rotation angle $\varphi$ of nanofins is set as the geometric parameter, ranging from 1° to 180° with a step size of 1°. For nanopillars, the radius $r$ of nanopillars is set as the geometric parameter, ranging from 25 nm to 125 nm with a step size of 1 nm. Notably, target meta-atom near the metasurface edges cannot be fully surrounded by five layers of neighboring meta-atoms in all directions. Consequently, simulation settings for the edge cases require separate treatment. Additional details are provided in Supporting Information S2. For nanopillar meta-atoms, only the $Ex$ component is characterized, as their symmetric structure exhibits polarization-insensitive behavior under $x$-polarized illumination, with cross-polarization conversion less than 2%. In contrast, for the asymmetric nanofin meta-atoms, which utilize PB phase modulation to control circular polarization, both $Ex$ and $Ey$ components must be measured simultaneously under circularly polarized illumination.

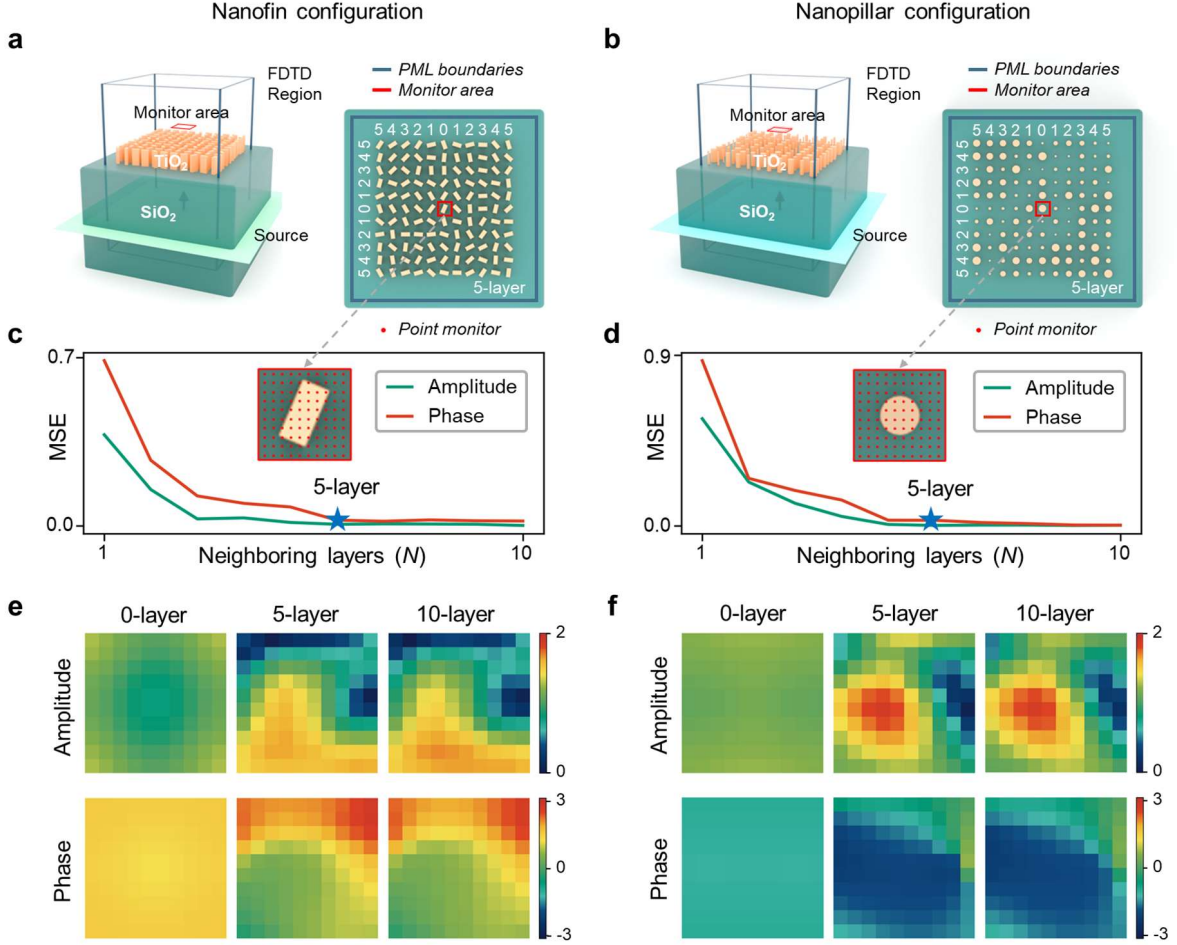

**Figure 2.** Simulation settings for nanofin configuration (left) and nanopillar configuration (right). (a-b) Schematic representation of the target TiO₂ meta-atoms, surrounded by randomly selected neighboring meta-atoms, positioned on a SiO₂ substrate. (c-d) MSE of the average amplitude and phase, sampled across a 10 × 10 array of point monitors, as functions of neighboring layers $N$. $N = 0$ represents the results derived with periodic boundaries. (e-f) Comparative amplitude and phase distributions of local responses for the selected 0-layer, 5-layer, and 10-layer models illustrating the evolution of local responses with increasing neighboring layers $N$.

## 3. Results

### 3.1 Forward Prediction

To efficiently predict the local responses of meta-atoms under various boundary conditions, a forward prediction network based on a multilayer perceptron (MLP) architecture is developed, as illustrated in **Figure 3**a. The input to the forward prediction network consists of the geometric parameters of the target meta-atom and those of its surrounding 5-layer neighboring meta-atoms. These geometric parameters ($\varphi$ for nanofin configuration and $r$ for nanopillar configuration) are vectorized into a 121 × 1 input vector $S_{input}$. To ensure robust training and numerical stability, the input vector is normalized before being fed into the MLP through two hidden layers containing 256 and 512 neurons, respectively. ReLU activation functions are applied after each hidden layer. Further training and test details are provided in Supporting Information S4. The output $\hat{R}$ of the forward prediction network consists of the real and imaginary parts of the local responses at target meta-atom's position, with dimensions of 400 × 1 for the $Ex$ and *Ey* nanofin configuration and 200 × 1 for the *Ex* nanopillar configuration. Using real and imaginary components avoids discontinuity at the $-\pi$ to $\pi$ phase transition, which is difficult for the network to learn, and instead provides two smooth, continuous quantities for training. The training dataset comprised over 10,000 samples, which were randomly divided into a training set (80%) and a test set (20%). During training, the MSE between the predicted local responses and the ground truth response (obtained via full-wave FDTD simulations) was employed as the loss function for training the MLP. To realize efficient full-wave-level prediction, we adopt a compact MLP rather than the deep, resource-intensive convolutional models used in prior studies. While the strong inductive biases of CNNs are powerful for learning spatial features from raw pixel data, they introduce significant computational overhead. We bypass this by explicitly encoding the spatial relationships into the input feature vector itself. The 2D structural parameter matrix of the meta-atoms and their neighbors is flattened into a 1D vector, where the data's ordering inherently preserves the neighboring information. Because this spatial topology is already encoded in the input, a lightweight MLP is sufficient to learn the mapping, greatly improving the efficiency.

As shown in Figure 3b, after 200 epochs of training, the average loss was $3.18 \times 10^{-2}$ for the training set and $3.54 \times 10^{-2}$ for the test set for nanofin configuration, and $1.51 \times 10^{-2}$ for the training set and $1.77 \times 10^{-2}$ for the test set for nanopillar configuration, respectively. Visualizations of local amplitude and phase responses predicted by the forward prediction network for three representative test samples are shown in Figure 3c–d. Although the phase distributions exhibit apparent discrepancies in certain regions (white dashed circles, Figure 3c–d, these variations primarily arise from the inherent periodicity of phase values, where $-\pi$ and $\pi$ represent identical physical states despite their numerical difference.

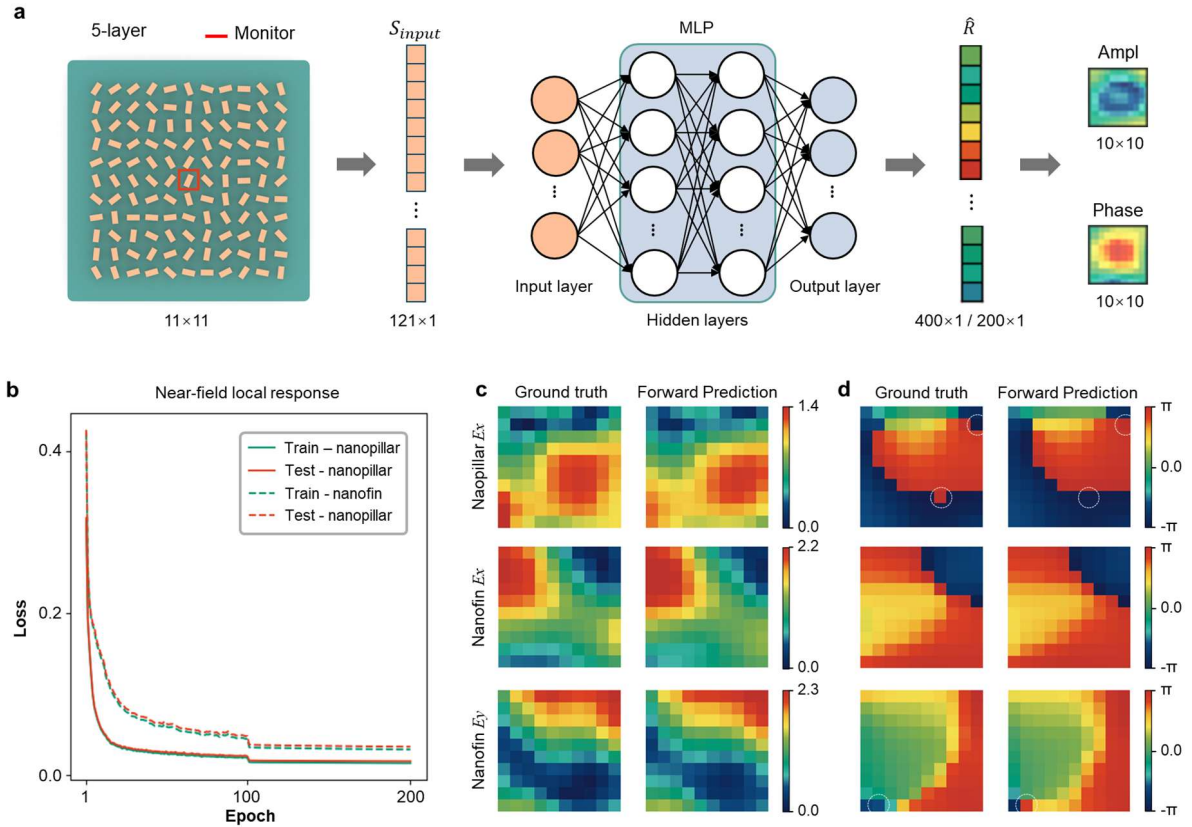

**Figure 3.** Forward prediction for local responses of target meta-atom. (a) Forward prediction network architecture. The input consists of the geometric parameters of a 5-layer model $S_{input}$. This input is then processed by a MLP network. The output $\hat{R}$ is the predicted real and imaginary parts of the local responses of the target meta-atom (highlighted by the red box). (b) Loss of the training set (green) and test set (orange) for both nanofin and nanopillar configurations during training. (c) Comparison between ground truth and forward-predicted amplitude distributions. (d) Comparison between ground truth and forward-predicted phase distributions.

By modeling the local response of a target meta-atom as a function of its own and neighboring geometric parameters, the forward prediction network effectively captures the mutual coupling effects within a localized region. This locality enables a scalable strategy, as illustrated in Figure 1, for predicting the near-field EM response of large-scale metasurfaces: an arbitrary $n \times n$ metasurface can be decomposed into $n^2$ overlapping 5-layer local patches, each centered on one target meta-atom. These sub-models are then independently processed by the forward prediction network, allowing the complete near-field EM response of the full metasurface to be rapidly reconstructed with full-wave-level accuracy. Compared to FDTD solvers, the forward prediction network achieved a 5,000-fold acceleration while requiring only 8 GB of memory. This characteristic enables large-scale metasurface inverse design by incorporating the pre-trained forward prediction network as a surrogate model for near-field response generation. **Table 1** summarizes the computational time required to predict the responses of metasurfaces with diameter ranging from micrometer (24 × 24 configuration) to centimeter (25,000 × 25,000 configuration). These results highlight the exceptional scalability of our framework, which can efficiently address metasurface sizes far beyond the capability of existing methods. In addition, compared with forward prediction models that account for coupling effects, our network delivers about 30-fold acceleration while reducing the prediction error to nearly one third of that reported in previous studies (see Supporting Information S1 for details). By providing fast and accurate EM predictions, the forward network supports the end-to-end training of inverse design networks on large and diverse datasets without relying on computationally expensive full-wave solvers.

**Table 1.** Computational time of forward prediction and inverse design of our framework

| Configuration | Prediction time | | Inverse design time | |
| --- | --- | --- | --- | --- |
| | nanopillar | nanofin | nanopillar | nanofin |
| 24 × 24 | 0.24 s | 0.22 s | 0.06 s | 0.06 s |
| 200 × 200 | 5.11 s | 5.93 s | 2.19 s | 2.26 s |
| 1,000 × 1,000 | 80.53 s | 108.49 s | 43.47 s | 45.23 s |
| 25,000 × 25,000 | 8 h 23 min | 9 h 47 min | 4 h 17 min | 4 h 58 min |

### 3.2 Inverse Design

To enable efficient large-scale inverse design of metasurfaces, a deep learning framework is developed, as illustrated in **Figure 4**, which integrates an inverse design module and a pre-trained forward prediction module. The inverse design module employs a U-Net architecture to reconstruct the metasurface structure $\hat{S}$ from the input near-field EM responses $R$[40]. To train this U-Net, a large dataset of paired examples ($R$, $S$) is required. We generate this dataset synthetically by leveraging the pre-trained forward network. We first generate a vast set of random metasurface structures S, and then compute their corresponding full-wave-accurate responses $R$ by applying the forward network to each S. This approach completely bypasses the need for computationally expensive FDTD simulations during dataset generation. The dimension of $R$ is (n, n, 400) for nanofin configuration and (n, n, 200) for nanopillar configuration, where the channels correspond to the real and imaginary components of the electric field. This approach ensures that the complete field information is maintained, since both the amplitude and phase of the optical field can be uniquely calculated from the real and imaginary parts. Given the full-wave accuracy achieved by the pre-trained forward prediction network, $R$ is synthesized by applying this network to a set of randomly generated metasurface structures. The inverse-designed metasurface structure $\hat{S}$ is represented as an n × n array, where each element corresponds to either the radius of a nanopillar or the rotation angle of a nanofin, depending on the metasurface type. This inverse mapping from field response to structural parameters allows the inverse design module to serve as the core component in determining the geometric parameters required to achieve the desired EM response. The forward prediction module consists of multiple MLP-based sub-networks proposed in the above section, each trained to approximate the local responses $\hat{R}$ of a target meta-atom by incorporating its geometric parameters and the influence of neighboring meta-atoms. During training, only the parameters of the reverse design module are updated, and the parameters of forward prediction module remain fixed, as it has been pre-trained separately. This ensures that the predicted metasurface structures yield physically consistent responses that remain consistent with full-wave simulation results. The MSE between the predicted near-field EM responses $\hat{R}$ and the input near-field EM response $R$ is chosen as the loss function to optimize the parameters of the inverse design module. Further training details, including data generation and implementation specifics, are provided in Supporting Information S3 and S4. A major advantage of this framework lies in its scalability. The inverse design module is capable of handling metasurfaces ranging from micrometer (24 × 24 configuration) to centimeter (25,000 × 25,000 configuration) in a single step without any iterative process, as shown in Table 1.

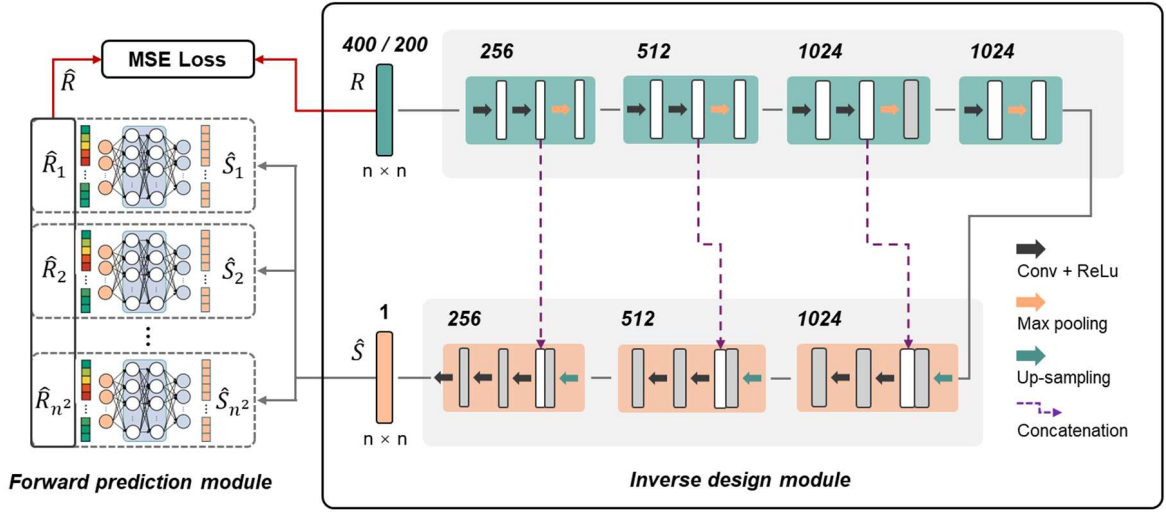

**Figure 4.** Schematic of the inverse design framework. The framework combines an inverse design module and a forward prediction module. Training is performed by minimizing the MSE between the predicted and target responses.

A comprehensive comparative analysis is conducted for the 24 × 24 configuration, including (a) inverse-designed metasurface $\hat{S}$ validated through FDTD simulations, (b) forward predictions of reference metasurface $S$, and (c) FDTD-computed ground truth of ref $S$. Amplitude and phase distributions are systematically compared across both near-field and Fresnel-field regions to evaluate the full-wave accuracy of the proposed methodologies. The forward-predicted results and the inverse-designed results are respectively compared against ground truth responses computed by FDTD simulations. As shown in Figure 5-7, both the forward-predicted and inverse-designed amplitude and phase distributions exhibit excellent agreement with the ground truth results. Apparent discrepancies observed in the phase distributions (white dashed circles in **Figure 5–7**) primarily result from the inherent periodicity of phase values, as discussed in forward prediction section. Quantitative comparisons of both inverse-designed and forward-predicted results against simulated near-field and Fresnel-region results are conducted using energy-normalized mean squared error (EN-MSE), defined as Equation 3:

$$EN - MSE = \frac{\sum_{ij}(\hat{A}[i,j] - A[i,j])^2}{\sum_{ij} A[i,j]^2} \qquad (3)$$

where $A[i,j]$ denotes the ground truth amplitude distribution and $\hat{A}[i,j]$ represents the predicted or inverse-designed amplitude distribution. For all tested configurations, both the inverse design and forward prediction results exhibit EN-MSEs below 3% in both near-field and Fresnel-field regions, demonstrating full-wave-level accuracy, as summarized in Table 2. Additional results and configuration-specific details of the inverse design performance can be found in Supporting Information S6 and S7. To further demonstrate the applicability of the proposed inverse design framework under broadband conditions, we performed additional inverse design tasks across multiple wavelengths ranging from 460 to 520 nm. This extended validation confirms the framework's robustness beyond single-wavelength operation. Related methods and results are provided the Supporting Information S8.

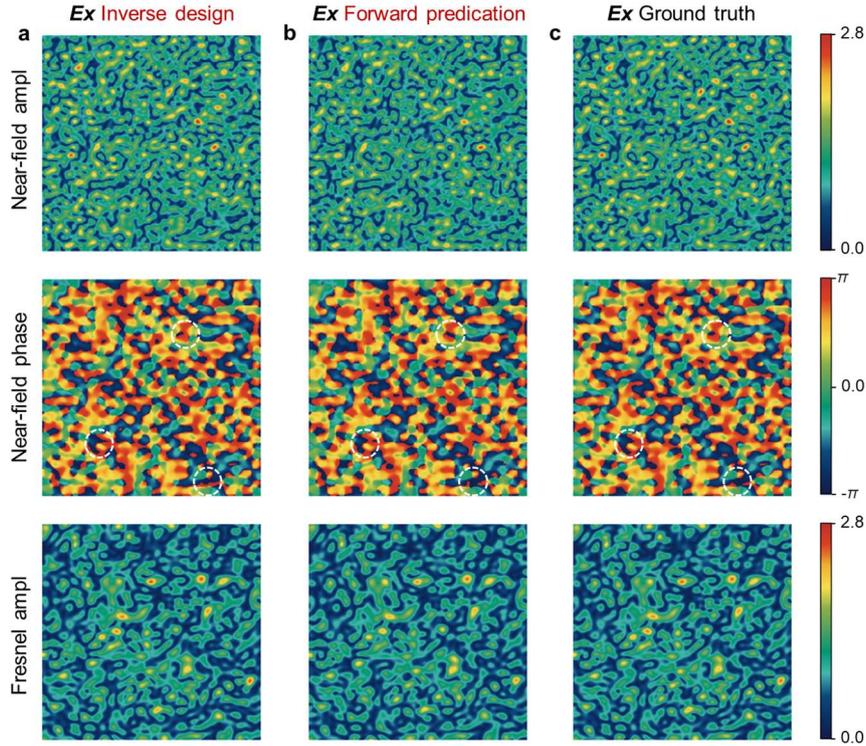

**Figure 5.** Comparison of *Ex* field distributions obtained via inverse design and forward prediction for a 24 × 24 nanofin configuration, with FDTD-computed results provided as ground truth in both near-field and Fresnel-field regions. White dashed circles represent the discrepancies arising from the inherent periodicity of phase values.

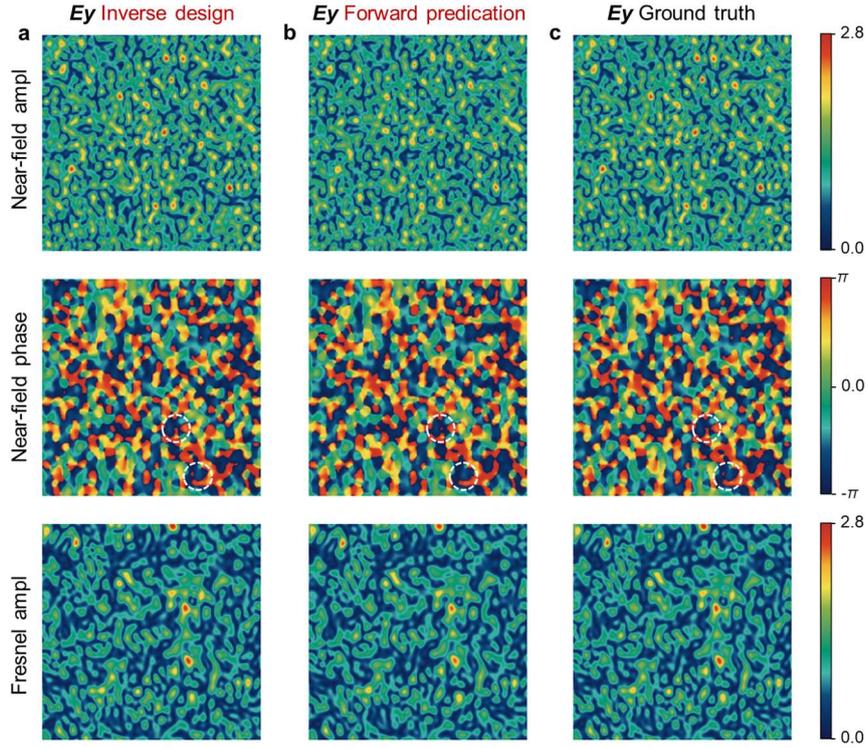

**Figure 6.** Comparison of *Ey* field distributions obtained via inverse design and forward prediction for a 24 × 24 nanofin configuration, with FDTD-computed results provided as ground truth in both near-field and Fresnel-field regions. White dashed circles represent the discrepancies arising from the inherent periodicity of phase values.

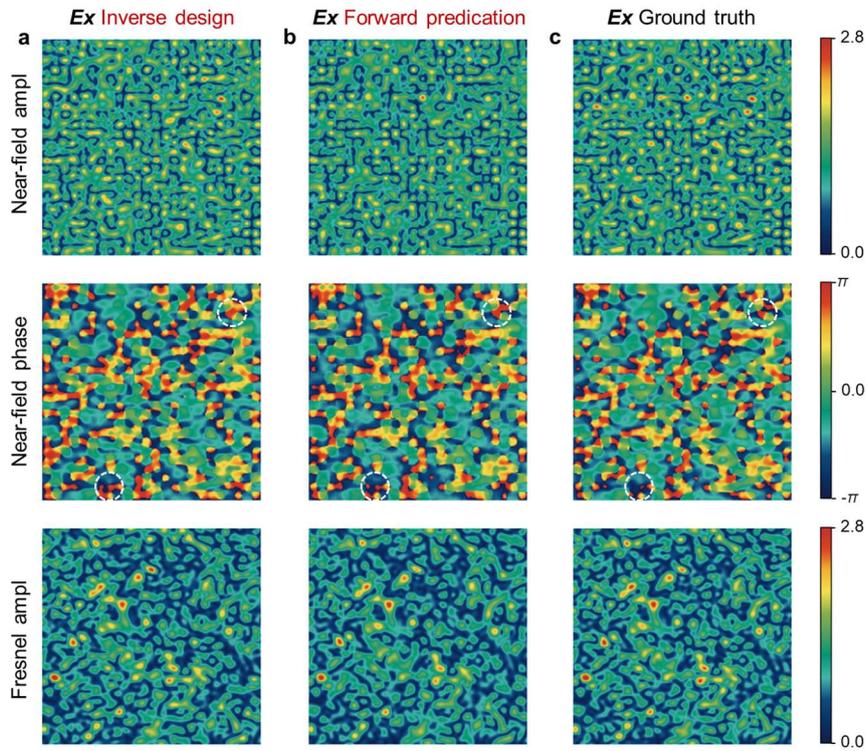

**Figure 7.** Comparison of *Ex* field distributions obtained via inverse design and forward prediction for a 24 × 24 nanopillar configuration, with FDTD-computed results provided as ground truth in both near-field and Fresnel-field regions. White dashed circles represent the discrepancies arising from the inherent periodicity of phase values.

**Table 2.** Comparison of energy-normalized mean squared error (EN-MSE) values for inverse design and forward prediction versus FDTD-computed results across both near-field and Fresnel-field regions.

| 24 × 24 Configuration | Inverse design | | Forward prediction | |
|---|---|---|---|---|
| | Near field | Fresnel field | Near field | Fresnel field |
| *Ex* nanofin **Figure 5** | 1.36% | 1.21% | 2.33% | 1.13% |
| *Ey* nanofin **Figure 6** | 2.45% | 2.21% | 2.36% | 1.19% |
| *Ex* nanopillar **Figure 7** | 1.70% | 1.79% | 2.53% | 2.78% |

### 3.3 Experimental Verification on Multi-foci Metalenses and Holographic Metasurfaces

To experimentally validate our inverse design framework, we designed, fabricated, and characterized four representative functional metasurfaces, as detailed in **Figure 8**. These devices were inverse-designed by our framework to match target near-field EM responses and subsequently fabricated using electron-beam lithography (EBL). We implemented both nanopillar and nanofin configurations, with each platform successfully demonstrating two distinct functionalities: multi-foci lensing and holographic projection. More design details are available in Supporting Information S9.

First, we demonstrated multi-foci lensing by fabricating two 224 × 224-pixel (89.6 μm × 89.6 μm) metasurfaces, one nanopillar-based and one nanofin-based. Both were designed to generate 12 discrete focal spots, uniformly distributed on a circle with 20 μm radius at a focal distance of $f$ = 0.1 mm. Second, for holographic projection, we fabricated two larger 560 × 560-pixel (224 μm × 224 μm) metasurfaces. The nanopillar-based device was optimized to reconstruct a holographic image of the letters "SJ", while its nanofin-based counterpart was designed to project "TU". Both images were designed to form at a propagation distance of $z$ = 0.5 mm.

Experimental characterization of all four fabricated samples showed excellent agreement with target simulation results. Optical measurements of both multi-foci metalenses confirmed that the 12 focal spots were accurately formed at the target plane of $f$ = 0.1 mm and precisely distributed on the circle with 20 μm radius, matching the design objectives. Furthermore, the holographic images "SJ" and "TU" were captured with high fidelity. To quantitatively evaluate the design accuracy, we employ two widely used full-reference metrics: the peak signal-to-noise ratio (PSNR) and the structural similarity index measure (SSIM)[41]. Quantitative comparison between the experimental results and the target simulations yielded strong agreement: the "SJ" hologram achieved PSNR of 24.19 dB and SSIM of 0.81, while the "TU" hologram achieved PSNR of 27.56 dB and SSIM of 0.82.

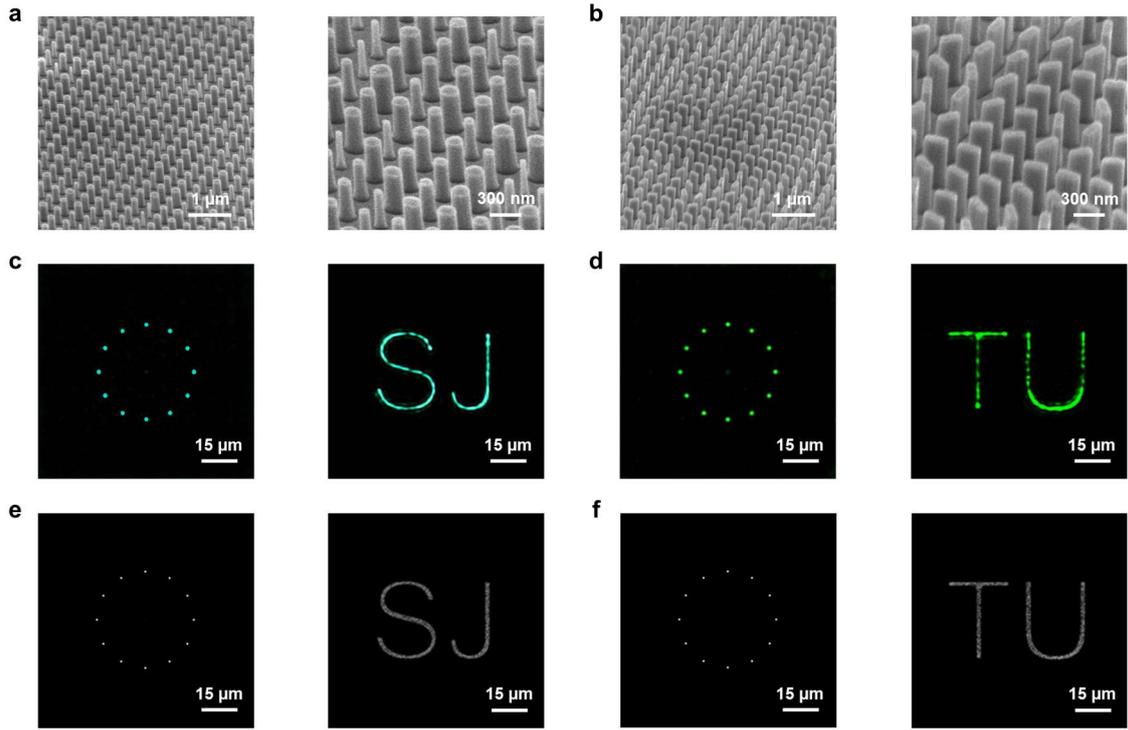

**Figure 8.** Experimental validation of the inverse-designed nanopillar and nanofin metasurfaces. (a) Scanning Electron Microscope (SEM) images of the fabricated nanopillar configuration in a titled view. (b) SEM images of the fabricated nanofin configuration in a titled view. (c) Experimentally measured intensity profiles for the nanopillar-based devices: (left) the multi-foci metalens and (right) the "SJ" hologram. (d) Experimentally measured intensity profiles for the nanofin-based devices: (left) the multi-foci metalens and (right) the "TU" hologram. (e) Corresponding target simulation patterns for the nanopillar devices (metalens and "SJ"). (f) Corresponding target simulation patterns for the nanofin devices (metalens and "TU").

## 4. Conclusion

In this work, we developed a general deep-learning-based inverse design framework for scalable and full-wave-accurate metasurface design. The proposed framework enables efficient design of aperiodic metasurfaces with extreme scale—ranging from micrometer to centimeter—by decomposing large metasurfaces into localized sub-models and leveraging a pre-trained forward prediction network. The framework not only works for the demonstrated scale range, but also is generalizable to metasurfaces of arbitrary size. It achieves near-field responses discrepancies blelow 3% compared to full-wave solvers at subwavelength ($< \lambda/10$) resolution. We experimentally demonstrated the effectiveness of the proposed framework through two representative applications: multi-foci metalenses and holographic metasurfaces. The fabricated multi-foci metalenses accurately generated the twelve designed focal spots, while the holographic metasurfaces successfully reconstructed the target images with high fidelity. The strong consistency between the experimental measurements and full-wave simulations confirms the robustness and reliability of the inverse design framework.

The inverse design network operates in a single forward pass and avoids computationally expensive iterative solvers, completing macroscale metasurface ($> 1k\lambda$) design tasks within tens of seconds using a standard four-core CPU. This dramatically improves the practicality of inverse design for real-world systems and opens the door to mass deployment of high-performance, functionally complex meta-optics. In the future, this inverse design paradigm can be extended to more complex scenarios involving multi-layered or multifunctional metasurfaces. Combining this framework with richer meta-atom libraries, such as bianisotropic or high-Q resonant structures, may further enhance amplitude and phase controllability of designed metasurfaces. As metasurfaces continue to push the limits of optical integration, imaging, and information processing, the demand for fast, scalable, and physically consistent design methods will continuously grow. We envision that the proposed framework, and future developments building upon such framework, will contribute significantly to the advancement of programmable, intelligent, and application-specific meta-optical platforms.

## 5. Methods

**FDTD Simulations.** Lumerical's FDTD solver is used to analyze mutual coupling effects and generating datasets for training forward prediction network. In the simulation setup, the height of the TiO2 nanofin and nanopillar meta-atoms is 500 nm, the periodicity of the unit cell is 400 nm, and the substrate is $SiO_2$. The nanofin and nanopillar configurations are illuminated under normal incidence with LCP and $x$-polarization, respectively. Point monitors are positioned 100 nm above the meta-atoms to capture near-field responses.

**Angular Spectrum Method.** To calculate the EM responses in Fresnel field, the angular spectrum method is utilized based on near-field. The process begins with the near-field EM responses obtained from the forward prediction network or FDTD simulations. The near-field data is then transformed into the Fresnel-field region ($z = 30$ μm) and the complex EM responses $U(x, y, z)$ at the Fresnel-field plane are given by the equation. More details of this method are provided in Supporting Information S5.

**Sample Fabrication.** The TiO2 metasurfaces were fabricated on double-side polished fused silica substrates. A 500-nm-thick TiO2 film was first deposited by ion-assisted deposition. The designed nanopillar patterns were defined by electron-beam lithography (EBL) on a positive resist layer, followed by development and the formation of a chromium (Cr) hard mask through a lift-off process. The patterns were then transferred into the TiO2 layer by inductively coupled plasma (ICP) etching using a fluorine-based gas mixture, and the Cr mask was subsequently removed to obtain the final TiO2 nanopillar and nanofin metasurface. Details of the fabrication process are provided in Section S10 of Supporting Information.

**Optical Characterization.** A supercontinuum laser was used to generate 500 nm and 532 nm light for illuminating the fabricated multi-foci metalenses and holographic metasurfaces. The unpolarized output beam was expanded, collimated, and linearly polarized before being incident on the samples. For nanopillar-based samples, the transmitted intensity profiles were captured using a 10× objective lens and a CMOS camera. For nanofin-based samples, the incident beam was converted into circularly polarized light by a quarter-wave plate (QWP) and followed the same optical path. Further details of the measurement and characterization setup are provided in Section S11 of Supporting Information.

## Acknowledgements

The work is financially supported by the National Natural Science Foundation of China (62375170, 22225501, W2412102, and 52421006), Shanghai Jiao Tong University under Grant YG2024QNA51, the Science and Technology Commission of Shanghai Municipality under Grant 20DZ2220400. Y. Mai also appreciates the Shanghai Municipal Science and Technology Major Project and SJTU 2030B plan (WH510207202).

**Data Availability Statement**

The data and codes that support the findings of this study are available upon reasonable request.

**Supporting Information**

Supporting information is available from the Wiley Online Library or from the author.

# Supporting Information

## S1: Comparison with State-of-the-Art Metasurface Design Frameworks

Recent advances have demonstrated the potential of deep learning in forward prediction and inverse design of metasurfaces. However, when applied to large-scale metasurfaces, these approaches often encounter challenges in scalability and computational efficiency, limiting their use for very large-area devices. In comparison, the inverse design framework presented in Fig. 4 of the main text achieves efficient performance for both forward prediction and inverse design, thereby overcoming the scale limitations and computational inefficiencies of the earlier approaches.

In forward prediction, many existing approaches neglect the electromagnetic coupling effects between adjacent structural units, and thus can only accurately predict the near-field response of isolated meta-atoms rather than that of full metasurfaces[1,2]. Methods that explicitly consider coupling effects often require excessively high sampling resolution, which makes full-wave scale predictions impractical[3], or employ prediction networks that are not efficient enough for large-scale metasurface inverse design[4]. In most cases, convolutional neural networks are adopted, leading to substantially larger model sizes compared with the lightweight MLP used in this work. For example, a recent graph convolutional network model[4] requires 43.9 s to predict the near-field response of a 10×10 metasurface configuration, whereas our MLP-based model accomplishes the same task in less than 1.5 s, corresponding to an approximate 30-fold acceleration. Under the same error metric, defined as $\frac{\|I_{Pred}-I_{FDTD}\|_2}{\|I_{FDTD}\|_2}$, our model achieves a maximum error of 0.11, representing a 2.6-fold improvement over the 0.29 error of the graph convolutional model.

Furthermore, most prior works rely on iterative optimization for inverse design. If 100 iterations are required, the graph convolutional approach would need more than three orders of magnitude longer runtime compared with the direct and efficient inverse design strategy proposed in this study. A detailed comparison of key parameters between our method and related works is further summarized in Table S1. Although previous studies have reported that graph convolutional networks can enable inverse design of large metasurfaces, the demonstrated maximum configuration size remains limited to 98×98 structural units[4], which is far smaller than the centimeter-scale metasurfaces of up to 25,000 × 25,000 units addressed in this work.

Table S1. Comparison of our method with existing deep-learning-enabled forward prediction and inverse design approaches for metasurfaces.

| Method | Wavelength (λ) | Coupling effects | Prediction network architecture | Sampling resolution | Inverse design method | Parameter count of prediction network |
|---|---|---|---|---|---|---|
| Nano Lett. (2020)[1] | 700 nm | Neglected | 91 layers with 33 Convs | <0.1λ | \ | ≈5.1 million |
| Nano Lett. (2024)[2] | 488 nm, 520 nm | Neglected | 5 Conv layes | <0.1λ | Iterative | ≈0.05 million |
| Adv. Opt. Mater. (2022)[3] | 545 nm, 1550 nm | Considered | 6 Conv layers | ≈0.5λ | Iterative | ≈1.4 million |
| ACS Photonics (2023)[4] | 600 nm | Considered | 6 Graph Conv layers | <0.1λ | Iterative | ≈300 million |
| This work | 500 nm, 520 nm | Considered | 4-layer MLP | <0.1λ | One-step | ≈0.3 million |

## S2: Generation of datasets for forward prediction

While the 5-layer model serves as the standard input configuration for most meta-atoms in a large metasurface, it does not apply to those located near the boundaries. Specifically, for an $n \times n$ metasurface, only the inner $(n - 10)^2$ meta-atoms are fully enclosed by five layers of neighboring structures. In contrast, edge-region meta-atoms are partially exposed to the surrounding environment and thus lack complete neighborhoods. To ensure accurate and consistent prediction of electromagnetic (EM) responses across the entire metasurface, these boundary cases must also be systematically addressed. To account for this, all possible incomplete neighborhood configurations are categorized. For the nanopillar configuration, which is polarization-insensitive, 36 representative neighborhood patterns are sufficient due to the presence of mirror symmetry. In contrast, the polarization-dependent nanofin configuration requires 72 distinct cases, as symmetry cannot be similarly exploited. These configurations are illustrated in Fig. S1.

Accordingly, during dataset generation, we do not solely record the local responses of center meta-atoms under the ideal 5-layer condition. Instead, local responses corresponding to all 36 representative meta-atoms are extracted and stored, as shown in Fig. S2. This comprehensive treatment enables the forward prediction network to generalize effectively across both interior and edge regions, ensuring robust performance on metasurfaces of arbitrary size.

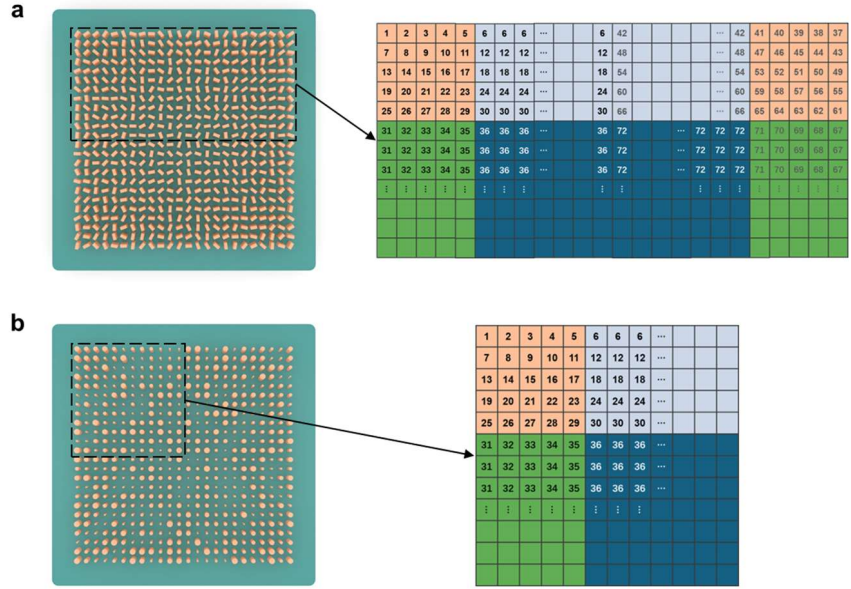

Fig. S1. Schematic of representative neighborhood configurations for edge-region meta-atoms in nanofin and nanopillar metasurfaces. (a) Nanofin configuration: 72 distinct neighborhood models are identified, accounting for polarization-dependent behavior and asymmetric surroundings. (b) Nanopillar configuration: 36 representative models are sufficient due to polarization insensitivity and mirror symmetry.

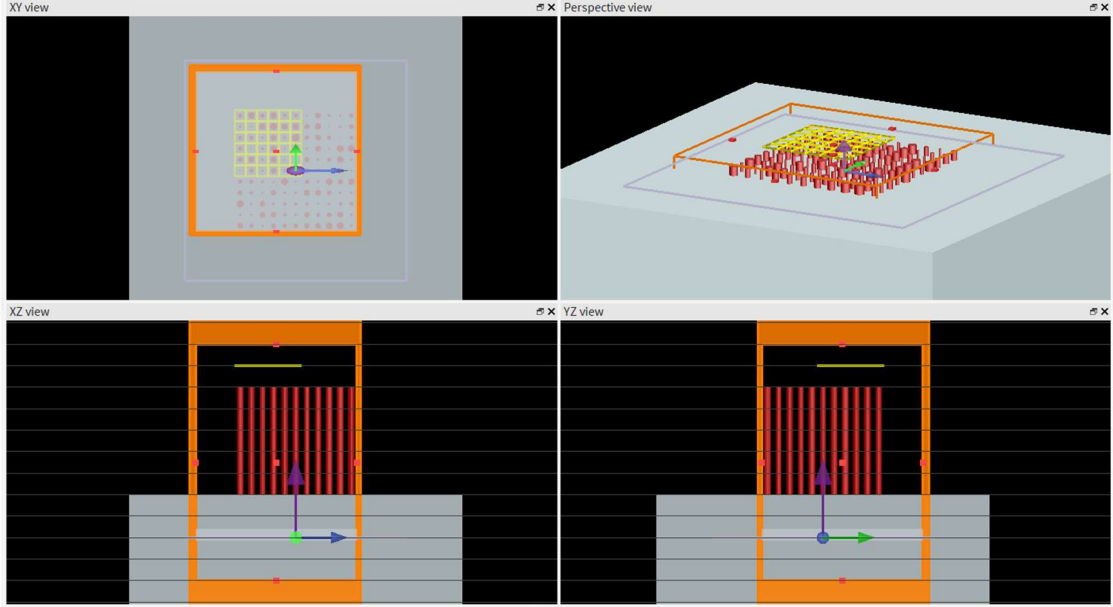

Fig. S2. Lumerical FDTD simulation setup for generating training and test data. For each metasurface, both interior and edge meta-atoms are considered.

## S3: Generation of datasets for inverse design

To construct the dataset for training the inverse design network, metasurface geometric parameters $S_{meta}$ were randomly generated for both nanofin and nanopillar configurations, as illustrated in Fig. S3. Each sampled $S_{meta}$ with dimensions of $n \times n$ was decomposed into $n^2$ local inputs $S_i$, which were then passed through the pre-trained forward prediction network to obtain the corresponding local response $R$. These responses served as inputs to the inverse design network. For the 24 × 24 configuration, 5,000 sets of $S_{meta}$ were randomly generated, while 500 sets were created for the 200 × 200 configuration. The resulting dataset was divided into a training set (80%) and a test set (20%) to ensure the network's generalization and performance on unseen data.

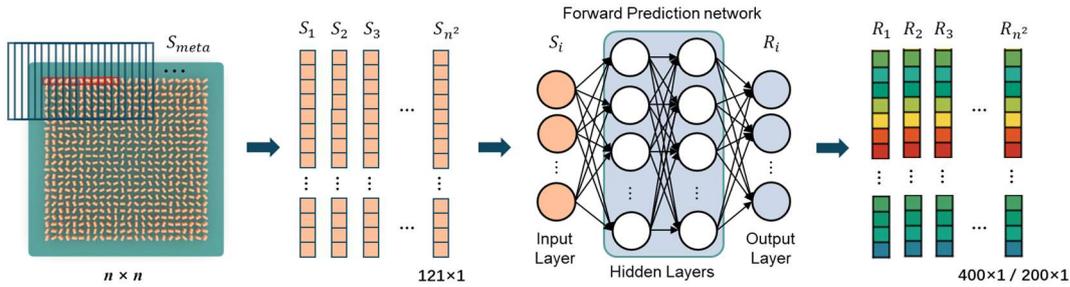

**Fig. S3. Schematic of the process for generating the dataset for the inverse design network, utilizing the pretrained forward prediction network.**

## S4: Training settings of forward prediction and inverse design networks

The forward prediction network and inverse design network training and testing settings are summarized in Table S2. All models were trained and evaluated on a system equipped with an NVIDIA GeForce RTX 4060 GPU (8 GB), an Intel i3-12100F CPU, and 32 GB of RAM. The networks were implemented in Python using the Pytorch 2.2.1 framework, while data pre-processing and post-processing were performed in MATLAB.

Table S2. Network training and test settings for three configurations.

| | Forward prediction | Inverse design | |
|---|---|---|---|
| | | 24 × 24 | 200 × 200 |
| Training set size | 8,000 | 4,000 | 400 |
| Test set size | 2,000 | 1,000 | 100 |
| Optimizer | Adam | Adam | Adam |
| Learning rate | 0.005 | 0.5 | 0.2 |
| Max iteration | 200 | 20 | 10 |
| Training time | ~2h | ~2h | ~18h |

## S5: Fresnel-field EM calculation using angular spectrum method

To calculate the Fresnel-field EM response, we use the angular spectrum method, which is based on the Fourier transform of the near-field EM distribution. This method is particularly effective for simulating the propagation of light through a metasurface and obtaining the field distribution at a distant imaging plane. The process begins with the near-field data obtained from the forward prediction network or FDTD simulations. The near-field data is then transformed into the Fresnel-field region using the angular spectrum approximation. Specifically, the complex electric field $U(x, y, z)$ at the Fresnel-field plane is given by:

$$A(f_X, f_Y; 0) = \iint U(x, y, 0) \exp[-j2\pi(f_X x + f_Y y)] dx dy \quad (S1)$$

$$A(f_X, f_Y; z) = A(f_x, f_y; 0) \exp\left(j2\pi z \sqrt{\frac{1}{\lambda^2} - f_X^2 - f_Y^2}\right) \quad (S2)$$

$$U(x, y, z) = \iint A(f_X, f_Y; z) \exp[j2\pi(f_X x + f_Y y)] df_X df_Y \quad (S3)$$

where $U(x, y, 0)$ represents the near-filed electric field, and $z$ is the distance to the observation plane, assuming that $f_X^2 + f_Y^2 < \frac{1}{\lambda^2}$, which is valid for our cases.

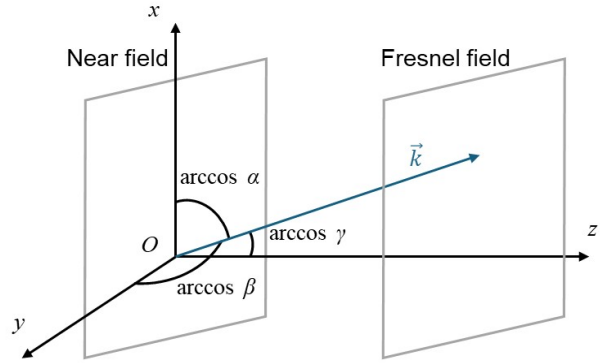

Fig. S4. Schematic of the angular spectrum method for transforming near-field EM data to the Fresnel-field response. The propagation direction of each spatial frequency component is described by the angles $\alpha$, $\beta$ and $\gamma$, where $\alpha = \lambda f_X$, $\beta = \lambda f_Y$, $\gamma = \sqrt{1 - \alpha^2 - \beta^2}$.

## S6: Local inverse design for large-scale metasurfaces

Due to the high computational demands associated with inverse design of ultra-large metasurfaces, such as a centimeter-scale (25,000 × 25,000) metasurface, we propose a local inverse design strategy as an alternative to global inverse design. The performance of this localized approach closely matches that of the global inverse design method described in inverse design section of the main text, with discrepancies in energy-normalized mean squared error (EN-MSE) below 1%. For instance, a 72 × 72 metasurface can be divided into a 3 × 3 array of smaller 24 × 24 sub-regions, as shown in Fig. S5. Each sub-region was independently inverse-designed using a neural network trained to map desired local responses to geometric parameters.

While this decomposition facilitates scalability, it introduces challenges near the boundaries of each sub-region, where neighboring predictions may lack continuity. The adjacent boundary conditions can be divided into 3 cases: corners, vertical edges and horizontal edges, as shown in Fig. S5. The primary inverse design network shown in Fig. 4 was used for the interior meta-atoms. The boundary regions were handled by auxiliary networks that have the same architecture as the primary network, excluding the output layer. As illustrated in Fig. S6, since boundary meta-atoms must remain compatible with those in the interior, they were generated from the auxiliary networks and then combined with the interior meta-atoms generated by the primary network to form a complete inverse-designed metasurface. The auxiliary networks were optimized by using the forward prediction module shown in Fig. 4 to compute the near-field response $\hat{R}$, which was then compared with the ground truth response $R$ through mean squared error minimization. As the influence of meta-atoms beyond a five-layer neighborhood is negligible for the near-field response of a target unit (see Fig. 2), the boundary dimensions were defined as $10 \times 10$, $(n-10) \times 10$ and $10 \times (n-10)$ for corners, vertical edges and horizontal edges, respectively. Moreover, because horizontal edges are equivalent to rotated vertical edges, a single auxiliary network was sufficient to account for both cases. The training strategy was kept identical to that of the primary network as shown in Table S2.

This primary–auxiliary coordination framework improves compatibility between adjacent sub-regions, reduces discontinuities in the predicted geometric parameters and ensures EM consistency across the entire metasurface. For instance, when a $72 \times 72$ metasurface is partitioned into nine $24 \times 24$ sub-regions, the resulting EN-MSEs for the nanofin $Ex$, nanofin $Ey$, and nanopillar $Ex$ fields are 0.33%, 0.33%, and 0.17%, respectively. The corresponding field distributions are presented in Fig. S7–S9.

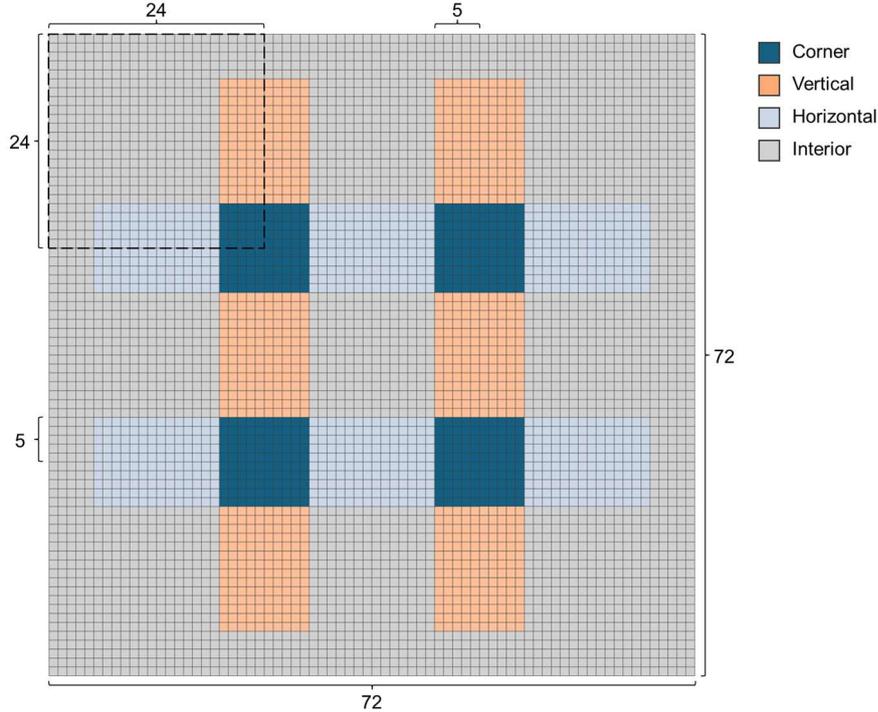

Fig. S5. Schematic of the local inverse design framework applied to a $72 \times 72$ nanofin metasurface. The metasurface is divided into nine $24 \times 24$ sub-regions (outlined by the dashed box), each designed independently. The adjacent boundary conditions can be divided into 3 cases: corners, vertical edges and horizontal edges.

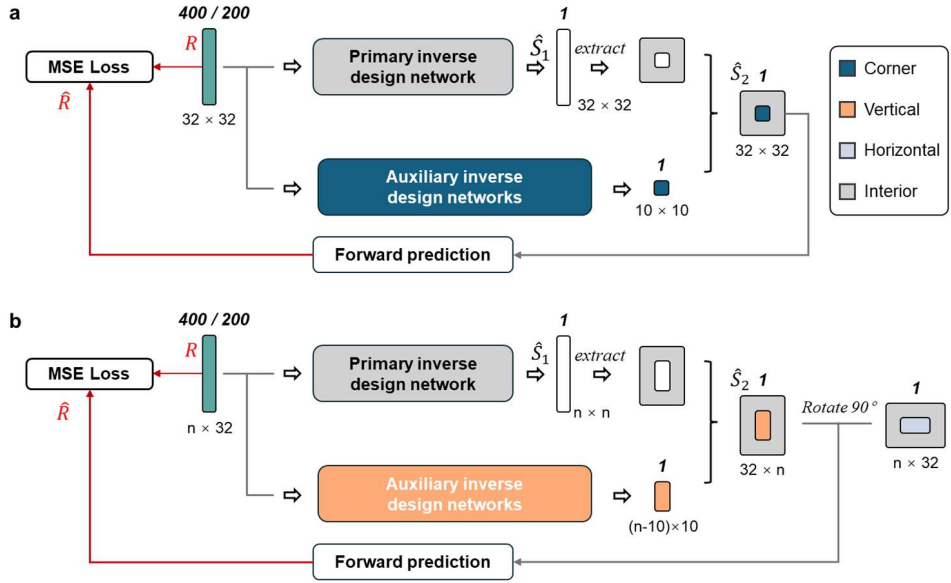

Fig. S6. Schematic of the inverse design framework with primary-auxiliary coordination for (a) corners and (b) vertical/horizontal edges.

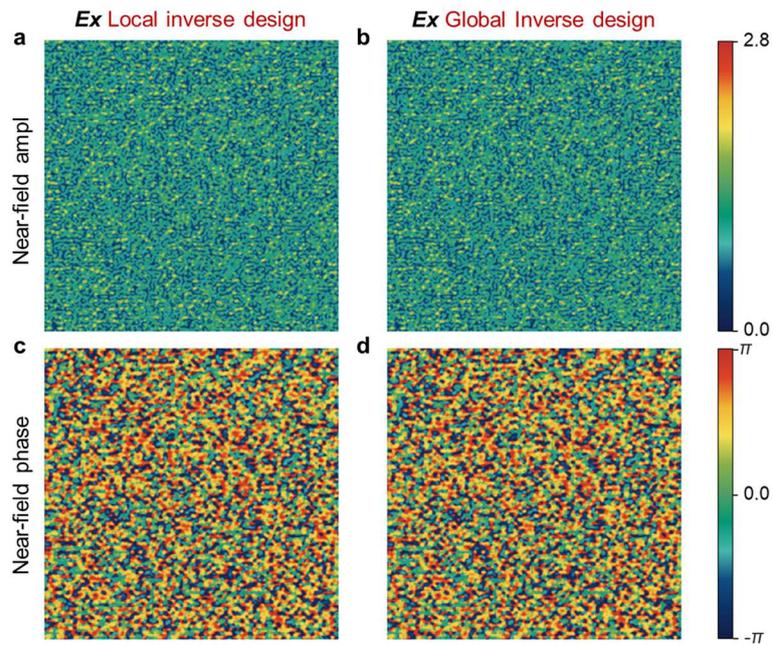

Fig. S7. Comparison of *Ex* field distributions obtained via local inverse design and global inverse design for a 72 × 72 nanofin configuration in the near-field region.

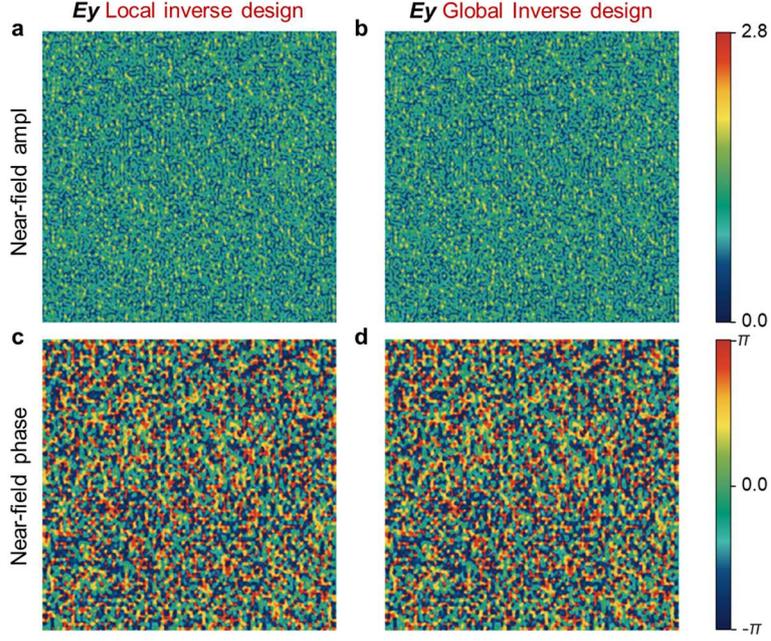

Fig. S8. Comparison of *Ey* field distributions obtained via local inverse design and global inverse design for a 72 × 72 nanofin configuration in the near-field region.

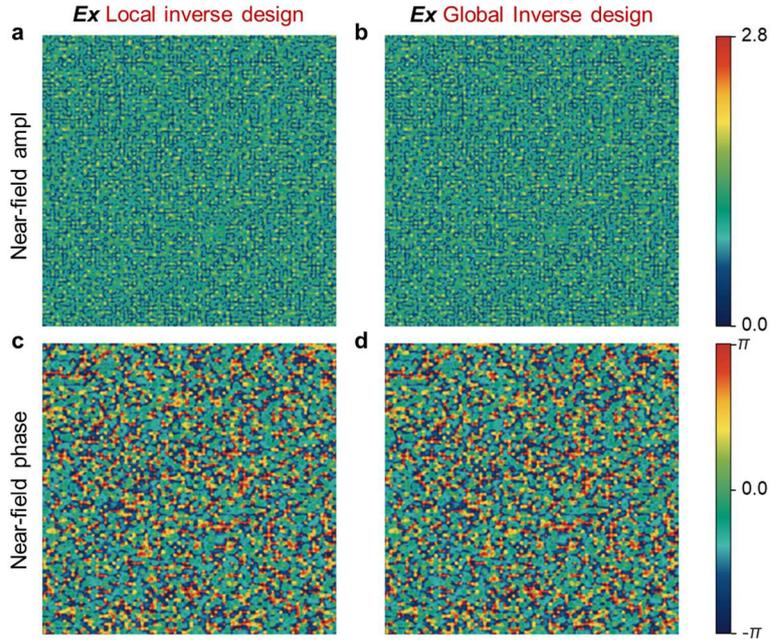

Fig. S9. Comparison of *Ex* field distributions obtained via local inverse design and global inverse design for a 72 × 72 nanopillar configuration in the near-field region.

## S7: Inverse design accuracy for varying scales

Table 2 in the main text summarizes the inverse design results compared with FDTD-computed results. However, for metasurfaces larger than 200 × 200, full-wave FDTD simulations become increasingly impractical due to the prohibitive computational resources required. To evaluate the scalability of our framework, we assess the accuracy of inverse design and forward prediction on metasurfaces with diameters ranging from the micrometer scale (24 × 24 configuration) to the centimeter scale (25,000 × 25,000 configuration), as reported in Table S3. Given that the EM responses predicted by our forward network demonstrate less than 3% deviation compared to FDTD simulations, it is reasonable to assess the accuracy of large-scale metasurface inverse design without full-wave simulations. Across all tested sizes, the EN-MSE remains below 2%, demonstrating that the proposed method maintains high accuracy even for extremely large-scale metasurface designs.

Table S3. Comparison of EN-MSE values between inverse design and forward prediction results in near-field region

| Configuration | 24 × 24 | 200 × 200 | 1,000 × 1,000 | 25,000 × 25,000 |
|---|---|---|---|---|
| $Ex$ nanofin | 0.20% | 0.16% | 0.33% | 0.56% |
| $Ey$ nanofin | 0.19% | 0.16% | 0.34% | 0.58% |
| $Ex$ nanopillar | 0.12% | 0.10% | 0.45% | 0.89% |

## S8: Broadband multi-wavelength inverse design

To extend the framework beyond monochromatic scenarios, we further performed broadband multi-wavelength forward prediction and inverse design tasks on a 24 × 24 nanopillar metasurface configuration at three representative wavelengths: 460 nm, 490 nm, and 520 nm. As shown in Fig. S10-S12, the inverse-designed amplitude and phase distributions exhibit excellent agreement with the ground truth results in both near-field and Fresnel-field regions. As reported in Table S4, the inverse design results achieved EN-MSE below 2% across all tested wavelengths. Interestingly, at 460 nm, the inverse design errors in both the near-field and Fresnel-field regions are slightly lower in the multi-wavelength case than in the single-wavelength case reported in Table 2. The improved performance under multi-wavelength settings may be attributed to the additional spectral constraints, which encourage the network to capture more generalizable structural-response relationships. By jointly optimizing across different wavelengths, the model is implicitly guided to learn deeper physical correlations such as dispersion effects and consistent geometric features, resulting in inverse design accuracy.

Table S4. Comparison of EN-MSE values between inverse design and ground truth across both near-field and Fresnel-field regions.

| $Ex$ | 460 nm | 490 nm | 520 nm |
|---|---|---|---|
| Near field | 1.65% | 0.88% | 0.73% |
| Fresnel field | 1.11% | 0.58% | 0.63% |

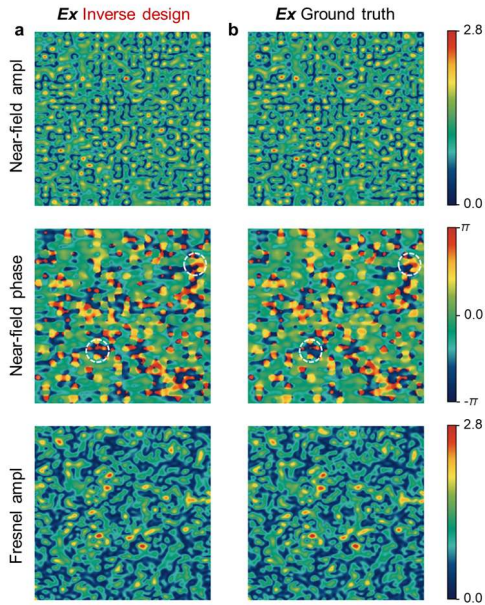

Fig. S10. Comparison of *Ex* ($\lambda$ = 460 nm) field distributions obtained from inverse-designed 24 × 24 nanopillar configuration and FDTD-computed ground truth, shown in both near-field and Fresnel-field regions.

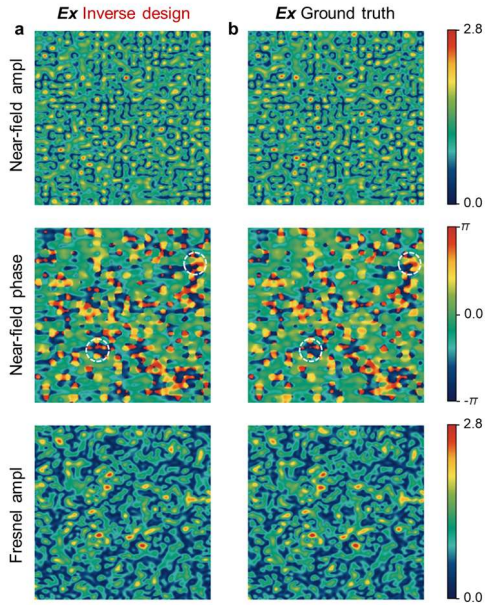

Fig. S11. Comparison of *Ex* ($\lambda$ = 490 nm) field distributions obtained from inverse-designed 24 × 24 nanopillar configuration and FDTD-computed ground truth, shown in both near-field and Fresnel-field regions.

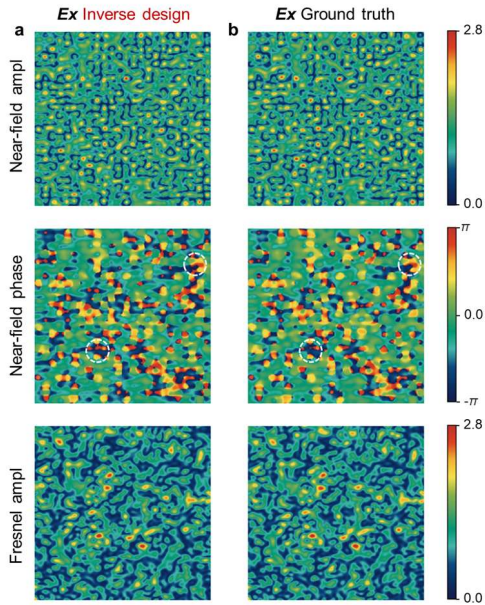

Fig. S12. Comparison of *Ex* ($\lambda$ = 520 nm) field distributions obtained from inverse-designed 24 × 24 nanopillar configuration and FDTD-computed ground truth, shown in both near-field and Fresnel-field regions.

## S9 Design method for multi-foci metalenses and holographic metasurfaces

A multi-foci metalens can be designed by coherently superposing the phase profiles of several single-focus metalenses. This approach enables the generation of distinct focal spots on a single focal plane at a distance. For a monochromatic plane wave, the phase profile required to generate a single focal point at coordinates is derived from:

$$\phi_j(x,y) = -\frac{2\pi}{\lambda_0}\left(\sqrt{f^2 + (x-x_j)^2 + (y-y_j)^2} - f_i\right), \quad f_j = \sqrt{f^2 + x_j^2 + y_j^2} \quad (S4)$$

The composite phase profile for the multi-foci metalens is then obtained through the complex superposition of these $N$ individual phase terms[5]:

$$\Phi_{total}(x,y) = arg\left(\sum_{j=1}^{N} e^{i\phi_j(x,y)}\right) \quad (S5)$$

This single, composite phase profile, when imparted to an incident plane wave, simultaneously focuses the light into off-axis spots on a single plane.

For a holographic metasurface, following the computer-generated holography framework demonstrated by Huang et al.[6], a holographic object can be modeled as an ensemble of discrete point sources. Each point source emits a spherical wavefront, and these waves interfere at the metasurface plane to form the desired complex field distribution. To emulate the diffuse reflection characteristics of a real object, a Gaussian-distributed random phase, $\delta_j$, is assigned to each source. The complex field contribution from the $j$-th point source at an arbitrary position $(x, y)$ on the metasurface is expressed as:

$$E_j(x,y) = \frac{A_j}{r_j} e^{i(kr_j + \delta_j)} \quad (S6)$$

where $A_j$ is the amplitude of the source, $k$ is the free-space wavenumber, $\delta_j$ is the random phase term, and $r_j$ is the distance from the point source to the position $(x_j, y_j)$ on the metasurface.

By coherently summing the contributions of all point sources, the total complex amplitude on the metasurface plane $E_{total}(x, y)$ is given by:

$$E_{total}(x,y) = \sum_j E_j(x,y) \quad (S7)$$

The target optical field of multi-foci metalenses and holographic metasurfaces is subsequently fed into the inverse-design neural network, which learns to map the desired field distribution to the corresponding local nanostructure parameters. By translating the target field into a physically realizable structural layout, the inverse-design framework generates a metasurface capable of accurately reconstructing the designed target under illumination.

## S10 Detailed fabrication process and parameters

The $TiO_2$ metasurfaces were fabricated on double-side polished fused silica substrates following a standard top-down nanofabrication procedure, as illustrated schematically in Fig. S13. A 500-nm-thick $TiO_2$ film was first deposited on the substrate by ion-assisted deposition (IAD). A positive electron-beam resist (ZEP520A) was then spin-coated on the $TiO_2$ surface and baked at 180 °C to remove residual solvent and enhance adhesion. The designed nanopillar and nanofin patterns were written into the resist by electron-beam lithography (EBL) operated at 100 kV acceleration voltage and 2 nA beam current. After exposure, the resist was developed in o-xylene and rinsed in isopropanol to form patterned openings.

A thin chromium (Cr) layer was subsequently deposited by electron-beam evaporation and patterned by a standard lift-off process to serve as a hard mask. The pattern was transferred into the $TiO_2$ layer using inductively coupled plasma (ICP) etching with a $C_4F_8/SF_6$ gas mixture optimized to achieve vertical sidewalls and minimal surface roughness. Finally, the Cr mask was removed by ICP dry etching, leaving free-standing $TiO_2$ nanopillar and nanofin arrays on the fused-silica substrate.

The fabricated metasurfaces were characterized by SEM to verify the geometry and uniformity of the nanopillars and nanofins, as presented in Fig. 8(a-b) of main text.

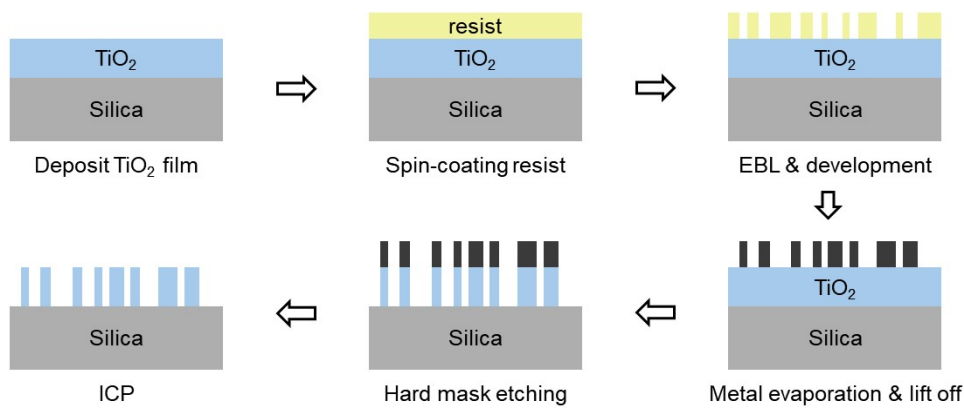

Fig. S13. Fabrication flow for the $TiO_2$-based metasurfaces.

## S11 Experiment setup of optical characterization

The optical measurement setup used for characterizing the multi-foci metalenses and holographic metasurfaces is schematically illustrated in Fig. S14. A supercontinuum laser source (SuperK EVO) was employed to generate broadband emission. Narrow spectral lines at 500 nm and 532 nm were selected from this source for use with nanopillar-based multi-foci metalenses and nanofin-based holographic metasurfaces, respectively. The unpolarized output beam from the laser was expanded and collimated through a beam expander, and subsequently passed through a linear polarizer (LP) to obtain a linearly polarized incident beam.

For nanopillar-based samples, the 500 nm linearly polarized beam was directly projected onto the sample surface. The transmitted light was collected by a 10× objective lens, collimated by a tube lens, and imaged onto a CMOS camera to record the transmitted intensity and focal spot distributions, as shown in Fig. S14(a). For nanofin-based samples, the 532 nm linearly polarized beam was converted into circularly polarized light by inserting a quarter-wave plate (QWP) after the linear polarizer. The circularly polarized beam was then directed onto the sample and followed the same imaging path as the nanopillar-based measurement configuration, as presented in Figs. S14(b). The captured images were used to verify the designed holographic functionalities.

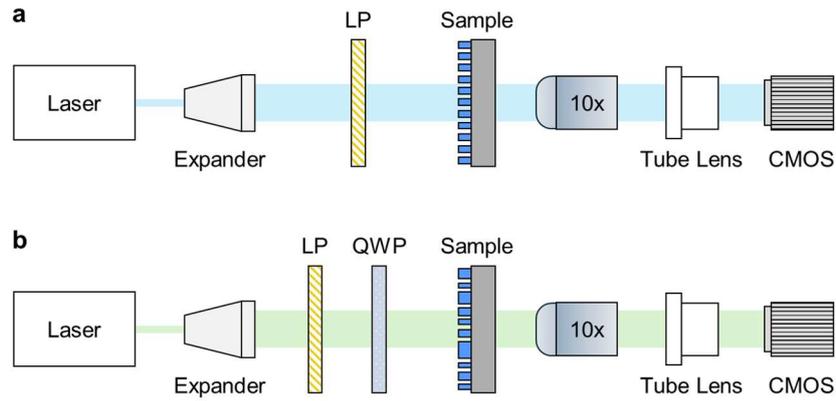

Fig. S14. Optical setup for experimentally verifying the performance of fabricated (a) nanopillar-based multi-foci metalenses and (b) nanofin-based holographic metasurfaces.